\documentclass{PoS}

\usepackage{amsfonts, amssymb}
\usepackage{amsmath, amsthm}
\usepackage{pstricks}
\usepackage{fontenc}
\usepackage{times}
\usepackage{mathptmx}
\usepackage{graphicx}
\usepackage{epsfig}

\title{Helium Compton Form Factor Measurements at CLAS}

\ShortTitle{He-DVCS@JLab.CLAS}

\author{\speaker{Eric VOUTIER} on behalf of the Jefferson Lab CLAS Collaboration\\
        \\
        Laboratoire de Physique Subatomique et de Cosmologie\\
	CNRS/IN2P3, Universit\'e Joseph Fourier, INPG\\
	53 avenue des Martyrs, 38026 Grenoble, France\\
	\\
        E-mail: \email{voutier@lpsc.in2p3.fr}}

\abstract{
The distribution of the parton content of nuclei, as encoded via the generalized parton distributions (GPDs), can be accessed via the deeply virtual Compton scattering (DVCS) process contributing to the cross section for leptoproduction of real photons. Similarly to the scattering of light by a material, DVCS provides information about the dynamics and the spatial structure of hadrons. The sensitivity of this process to the lepton beam polarization allows to single-out the DVCS amplitude in terms of Compton form factors that contain GPDs information. The beam spin asymmetry of the $^4$He($\vec {\mathrm e}$,e$' \gamma ^4$He) process was measured in the experimental Hall B of the Jefferson Laboratory to extract the real and imaginary parts of the twist-2 Compton form factor of the $^4$He nucleus. The experimental results reported here demonstrate the relevance of 
this method for such a goal, and suggest the dominance of the Bethe-Heitler amplitude to the unpolarized process in the kinematic range explored by the experiment. 
}

\FullConference{XXI International Workshop on Deep-Inelastic Scattering and Related Subject - DIS2013,\\
		22-26 April 2013\\
		Marseille,France}

\begin{document}

\section{Physics motivations}

Generalized parton distributions (GPDs)~\cite{Mul94} offer the unprecented possibility to access the spatial distribution of partons inside nucleons and nuclei~\cite{{Die03},{Bel05}}. Interpreted as the distribution in the transverse plane of partons carrying a certain longitunal momentum fraction, they allow for a tri-dimensional representation of the parton content of nuclear matter~\cite{Bur00}. Encoding the correlations between partons, GPDs also contain information about the dynamics of the system like the parton contribution to the angular momentum of the nucleon~\cite{Ji97}, or the distribution of strong forces experienced by partons inside hadrons~\cite{Pol03}. 

GPDs can be accessed experimentally via the exclusive deeply virtual Compton scattering (DVCS) reaction. DVCS off nuclear targets allow to address the partonic structure of the nucleus (coherent channel)~\cite{{Sco04},{Liu05}} as well as the partonic structure of nucleons (incoherent channel) embedded in the nuclear medium~\cite{Guz09}. Nuclear DVCS allows to investigate the role of the transverse degrees of freedom in the nuclear partonic structure, and is expected to be strongly sensitive~\cite{Liu05} to the specific origin of the EMC effect~\cite{Aub83}. The transverse sensitivity can theoretically be expressed in terms of the generalized EMC ratio, corresponding to 
the ratio of nuclear GPDs with free nucleon GPDs and leading to the usual EMC ratio in the forward limit. 

The high density of the $^4$He nucleus and its isoscalar nature have always been appealing features for nuclear structure studies. In the case of DVCS experiments, this translates into the ability to extract from the single measurement of the beam spin asymmetry (BSA), the real and imaginary parts of the Compton form factor that contains the $^4$He leading twist GPD. This presentation reports about the first pass analysis of the eg6 run at CLAS studying the He-DVCS coherent channel.

\section{Experimental method}

The DVCS reaction $^4$He($\vec {\mathrm e}$,e$' \gamma ^4$He) was measured~\cite{Egi08} during the eg6 run taking place in the experimental Hall B of the Jefferson Laboratory. A continuous electron beam of 6.065~GeV longitudinally polarized at -83.7\% interacted with a 20~cm long gazeous helium target pressurized at 5.1~atm and contained in a 3~mm radius kapton cylinder. The knowledge of the electron beam energy and the measurement of scattered electrons with the CLAS detector~\cite{Mec03} allow for a precise tagging of virtual photons probing the $^4$He nucleus. Real photons were detected in a small angle electromagnetic calorimeter (IC) specifically developed for DVCS experiments at CLAS~\cite{Gir08} and achieving during this experiment a 7.1\% resolution at the pion mass. Identification of the coherent DVCS channel requires the detection of the recoil $^4$He nucleus for a clear separation from quasi-elastic (incoherent) channels. This was accomplished with a cylindrical radial time projection chamber (RTPC) surrounding the reaction target and detecting recoil helium nucleus above $\sim$300~MeV/c. The eg6-RTPC is a mechanically modified version of the BoNuS TPC~\cite{Fen08} operating at variance with a Ne(80\%)-DME(20\%) gas mixture at room temperature and atmospheric pressure, thanks to the large stopping power of $^4$He particles. The RTPC is immersed in a strong solenoid field that shields the detector with respect to low energy M\o ller electrons, as well as the IC which is attached to the downstream end of the DVCS solenoid magnet.

\begin{figure}[t]
\begin{center}
\epsfig{file=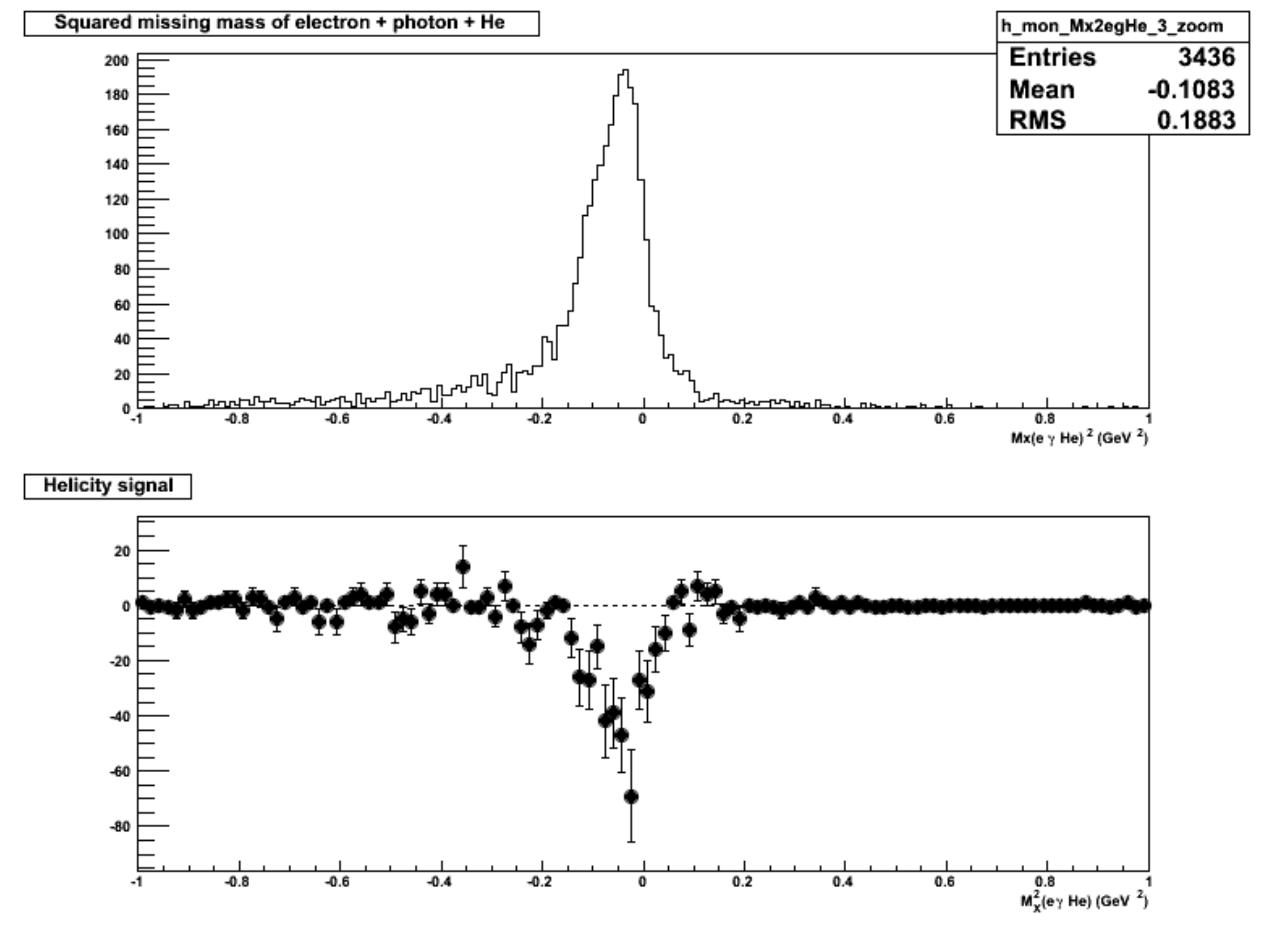, height=7.5cm}
\vspace*{-10pt}
\caption{Missing mass spectra of the process $^4$He(e,e$' \gamma ^4$He)X and corresponding helicity signal for the first pass analysis of eg6 data~\cite{Per12}.} \label{fig:mismas}
\end{center}
\end{figure}

The selection of coherent DVCS events relies on the particle identification capabilities of each detector and a set of physics criteria specific to the kinematics of the DVCS reaction. The RTPC response to $^4$He particle was calibrated with elastic scattering data where angle and momenta of recoil $^4$He are reconstructed from electron kinematics. This tagged $^4$He flux, in a similar energy range than $^4$He-DVCS, allows to calibrate the RTPC drift paths and provides a preliminar calibration of the energy deposit collected by each pad constituting the RTPC detection surface. Due to the  partial illumination of the RTPC resulting form the scattered electron detection, this technique is however limited and does not provide the full energy calibration necessary for the separation between the different recoil nucleus that could be produced. The selection of one single charged particle in CLAS and one single photon in IC helps to reduce the potential contamination of candidate $^4$He-DVCS events. The correlation between the  energy and the angle of real photons produced in coherent DVCS, and between the out-of-plane angles of real photons ($\varphi)$ and recoil $^4$He are the basic physics constraints applied to experimental data. Additional cuts on the transverse missing momentum and the squared missing mass (Fig.~\ref{fig:mismas}) of the process $^4$He(e,e$' \gamma ^4$He)X allow for the final selection of $^4$He-DVCS events.

The eg6 experiment measured the sensitivity of the DVCS reaction to the helicity of the electron beam. This can be expressed in terms of the helicty signal defined as 
\begin{equation}
S_h = \int_0^{\pi} \left( N^+(\varphi) - N^-(\varphi) \right) \, d\varphi - \int_{\pi}^{2\pi} \left( N^+(\varphi) - N^-(\varphi) \right) \, d\varphi \label{eq:hel}
\end{equation}
where $N^{\pm}$ denotes the number of events for the $\pm$ beam helicity. It is represented on Fig.~\ref{fig:mismas} for the full data set of the experiment within the first pass analysis. It is particularly seen that a strong helicity signal is obtained in the missing mass region corresponding to $^4$He-DVCS candidate events, enabling a significative extraction of the $^4$He Compton form factor. 

\section{Compton form factor extraction}

The real and imaginary parts of the twist-2 Compton form factor ${\mathcal H}_A$ of the $^4$He nucleus are obtained from 
the analysis of the out-ot-plane distribution of the beam spin asymmetry $A^{\lambda}_{LU}$ following the twist-2  expression~\cite{Bel01}
\begin{equation}
\hspace*{-3pt}
A^{\lambda}_{LU}(\varphi) = \frac{N^+(\varphi) - N^-(\varphi)}{N^+(\varphi) + N^-(\varphi)} = \frac{\lambda \, \alpha_0(\varphi) \Im m \{{\mathcal H}_A\}}{\alpha_1(\varphi) + \alpha_2(\varphi) \Re e \{{\mathcal H}_A\} + \alpha_3(\varphi) {\left[ {\Re e \{ {\mathcal H}_A \} }^2 + {\Im m \{{\mathcal H}_A\}}^2 \right]}^{\phantom{1}} } \label{eq:phi_fit}
\end{equation}
where $\alpha_i$'s are known $\varphi$-dependent kinematical coefficients reconstructed from the measured kinematics of selected events, and $\lambda$ represents the beam polarization. For the kinematical domain explored by this experiment ($-t<0.2\,\,\mathrm{GeV}^2$, $1.0\,\,\mathrm{GeV}^2<Q^2<2.3\,\,\mathrm{GeV}^2$, and $0.10<x_B<0.25$) the magnitude of $\alpha_i$ coefficients is $\alpha_0 \sim 10^{-2}$, $\alpha_1 \sim 1$, $\alpha_2 \sim 10^{-2}$, and $\alpha_3 \sim 10^{-4}$ 
suggesting the dominance of the Bethe-Heitler contribution to the unpolarized cross section and consequently a reduced sensitivity to the real part of the Compton form factor.  

\begin{figure}[t]
\begin{center}
\epsfig{file=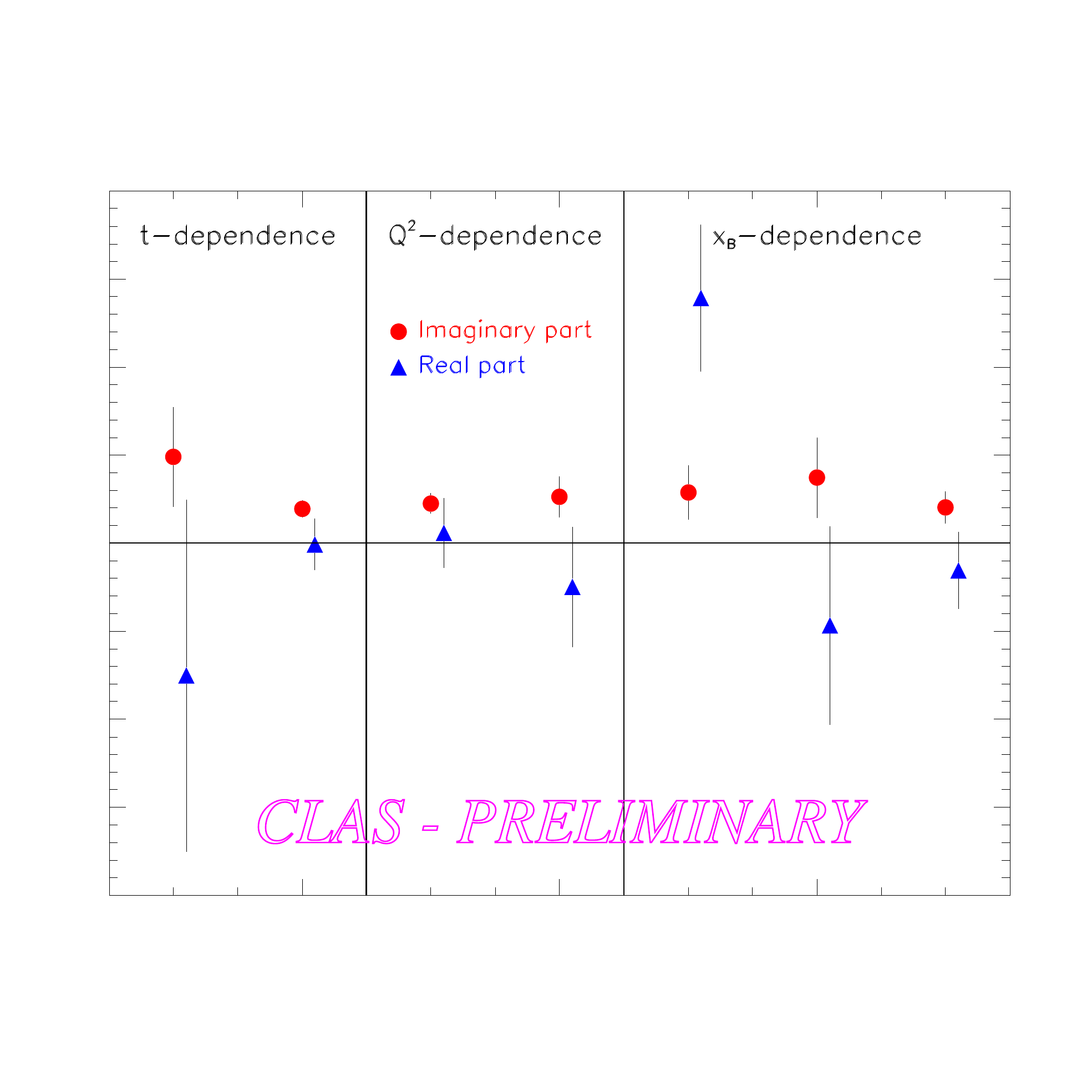, height=8.0cm}
\vspace*{-40pt}
\caption{Real and imaginary parts (arbitrary units) of the twist-2 $^4$He Compton form factor as extracted from the first pass analysis of $A^{\lambda}_{LU}$ measurements~\cite{Per12}; each panel represents the dependence with respect to one kinematical parameter while integrating events over the other parameters; data points are put on the same $y$-scale, with zero indicated by the horizontal 
axis, and sligthly shifted in $x$ for better reading; error bars are statistical only.} \label{fig:cff}
\end{center}
\end{figure}

The limited statistics of the experiment does not allow for a simultaneous multi-dimensional binning of the distribution of events in the DVCS variables $(t,Q^2,x_B,\varphi)$. Instead, a two-dimensio-nal binning in $\varphi$ and one of the other variable is applied to data: for example, $(t,\varphi)$ binning while integrating events over the $(Q^2,x_B)$ experimental phase space. This procedure allows to study the separate dependence of the Compton form factor on each physics kinematical parameter. For each bin, the average value of $\alpha_i$ coefficients was considered and the expression Eq.~\ref{eq:phi_fit} was fitted to data to extract the real and imaginary parts of the Compton form factor. Preliminary experimental results are shown on Fig.~\ref{fig:cff} in arbitrary units while real and imaginary parts are put on the same $y$-scale. As expected from the magnitude of $\alpha_i$ coefficients, a good determination of the imaginary part is obtained while the real part suffers from larger fluctuations originating essentially from lack of statistics. Nevertheless these results demonstrate the relevance of the BSA method for the experimental measurement of the real and imaginary parts of the twist-2 Compton form factor of the $^4$He nucleus. The on-going final analysis will provide the definitive experimental values of the $^4$He Compton form factor  featuring refined calibration of the CLAS detector and the full energy calibration of the RTPC. The latter is expected to remove possible contamination of $^4$He-DVCS candidate events from incoherent channels, and to allow for a precise determination of possible $\pi^0$ contamination of experimental data. 
 
\section{Conclusion}

The first pass analysis of coherent $^4$He-DVCS BSA experimental data acquired during the eg6 run at CLAS allows for the experimental determination of the real and imaginary parts of the twist-2 Compton form factor of the $^4$He nucleus. Data  analysis shows that the real part extraction is more sensitive than the imaginary part to the experiment statistics, as a consequence of the relative importance of the contribution of the real part to the unpolarized cross section. This suggests that, in the kinematic range of this experiment, the Bethe-Heitler process dominates the unpolarized cross section for electroproduction of real photons off $^4$He, at variance with previous observations off the nucleon~\cite{Mun06}.    

The results of this exploratory experiment set the basis of a dedicated experimental program for the study of nuclear DVCS 
at JLab 12~GeV, in the perspective of a tri-dimensional imaging of the partonic structure of nuclei and a better understanding of the EMC effect.

\end{document}